\documentclass[twocolumn]{aastex62}
\usepackage{float}
\usepackage[flushleft]{threeparttable}
\usepackage{footnotebackref}
\usepackage{xcolor}

\shorttitle{U~Mon}
\shortauthors{Vega et al. 2021}

\begin{document}

\title{\bf Multiwavelength Observations of the RV~Tauri Variable System U~Monocerotis: Long-Term Variability Phenomena That Can Be Explained by Binary Interactions with a Circumbinary Disk}

\author[0000-0002-5928-2685]{Laura D. Vega}
\altaffiliation{NASA MUREP Harriett G. Jenkins Predoctoral Fellow}
\altaffiliation{Fisk-Vanderbilt Master's-to-PhD Bridge Program}
\affiliation{Department of Physics \& Astronomy, Vanderbilt University, Nashville, TN 37235, USA}
\affiliation{Astrophysics Science Division, NASA Goddard Space Flight Center, Greenbelt, MD 20771, USA}

\author[0000-0002-3481-9052]{Keivan G. Stassun}
\affiliation{Department of Physics \& Astronomy, Vanderbilt University, Nashville, TN 37235, USA}

\author[0000-0002-6752-2909]{Rodolfo Montez Jr.}
\affiliation{Center for Astrophysics $\vert$\ Harvard\ \&\ Smithsonian, Cambridge, MA 02138, USA}

\author[0000-0001-8541-8024]{Tomasz Kami{\'n}ski}
\affiliation{Nicolaus Copernicus Astronomical Center of the Polish Academy of Sciences, Toru\'n, Poland}

\author[0000-0003-0242-0044]{Laurence Sabin}
\affil{Instituto de Astronom{\'i}a, Universidad Nacional Aut{\'o}noma de M{\'e}xico, Ensenada, B. C., M{\'e}xico}

\author[0000-0002-4162-8190]{Eric M. Schlegel}
\affiliation{Department of Physics \& Astronomy, The University of Texas at San Antonio, San Antonio, TX 78249, USA}

\author[0000-0002-2700-9916]{Wouter H. T. Vlemmings}
\affiliation{Department of Space, Earth and Environment, Chalmers University of Technology, Onsala Space Observatory, Onsala, Sweden}

\author[0000-0002-3138-8250]{Joel H. Kastner}
\affiliation{School of Physics and Astronomy, Rochester Institute of Technology, Rochester, NY 14623, USA}

\author[0000-0001-5177-6202]{Sofia Ramstedt}
\affiliation{Department of Physics and Astronomy, Uppsala University, Uppsala, Sweden}

\author[0000-0003-0442-4284]{Patricia T. Boyd}
\affiliation{Astrophysics Science Division, NASA Goddard Space Flight Center, Greenbelt, MD 20771, USA}

\begin{abstract}
We present X-ray through submillimeter observations of the classical RV~Tauri (RVb-type) variable U~Mon, a post-AGB binary with a circumbinary disk (CBD). Our SMA observations indicate a CBD diameter of $\lesssim$550~au. Our XMM-Newton observations make U~Mon the first RV~Tauri variable detected in X-rays. The X-ray emission is characteristic of a hot plasma ($\sim$10~MK), with L\textsubscript{X}$=5\times10^{30}{\rm ~erg}{\rm ~s}^{-1}$, and we consider its possible origin from U Mon, its companion, and/or binary system interactions. Combining DASCH and AAVSO data, we extend the time-series photometric baseline back to the late 1880s and find evidence that U~Mon has secular changes that appear to recur on a timescale of $\sim$60~yr, possibly caused by a feature in the CBD. From literature radial velocities we find that the binary companion is a $\sim$2~M$_{\odot}$ A-type main-sequence star. The orientation of the binary's orbit lies along our line of sight ($\omega = 95^\circ$), such that apastron corresponds to photometric RVb minima, consistent with the post-AGB star becoming obscured by the near side of the CBD. In addition, we find the size of the inner-CBD hole ($\sim$4.5--9~au) to be comparable to the binary separation, implying that one or both stars may interact with the CBD at apastron. The obscuration of the post-AGB star implicates the companion as the likely source of the enhanced H$\alpha$ observed at RVb minima and of the X-ray emission that may arise from accreted material.
\end{abstract}


\received{2020 August 3}
\revised{2021 January 29}
\accepted{to ApJ 2021 February 2}
\published{2021 March 12}

\section{Introduction} \label{sec:intro}

RV~Tauri-type variable stars are F through K supergiants with luminosity classes between Ia and II. They extend the brightest part of the Type II Cepheid (i.e., RR~Lyr, BL~Her, W~Vir, and RV~Tau stars) instability strip on the H-R diagram, as well as the brightest part of their period-luminosity relation, with radial pulsation periods longer than $20$~days \citep{Soszynski2017}. However, it has been shown that RV~Tauri variables follow a steeper period-luminosity relation compared to their shorter-period Type II Cepheid counterparts \citep[e.g.][]{Bodi2019}.

The signature characteristic of RV~Tauri variables are the alternating deep and shallow minima in their light curves: the time between a deep minimum and its successive shallower minimum is called the ``fundamental'' period; the time between two successive deep minima is the ``formal'' period. The alternation of deep and shallow minima is not strict in nature, as the light curves also show strong cycle-to-cycle variability, where the depths of the minima can vary randomly and may be attributed to low-dimensional chaotic behavior \citep[e.g.][and references therein]{Plachy2014,Plachy2018}. In addition to pulsations, a photometric subgroup (RVb-type) displays an additional, slower, periodic variation in mean brightness that ranges between 470 and 2800~days \citep{Soszynski2017}. There are $\sim$300 identified RV~Tauri variables in the Galaxy \citep{Soszynski2020}.

It has been established that most RV~Tauri variables are a subclass of post-asymptotic giant branch (post-AGB) stars in binary systems surrounded by a dusty circumbinary disk (CBD), and there is argument for the companion to be an unevolved main-sequence star \citep{Vanwinckel2009,Manick2019}. The post-AGB primary is evolved from a low-to-intermediate-mass progenitor (0.8--8~M$_{\odot}$), on their way to forming a planetary nebula, based on high luminosities ($\sim$10$^3$--$10^4$~L$_{\odot}$), mass-loss history, and infrared excess (indicative of dust) in their spectral energy distribution \citep[SED;][]{Gehrz1972,LloydEvans1985, Jura1986, Alcock1998,vanwinckel2003}. \citet{Groenewegen_Jurkovic2017} recently found mass estimates for RV~Tauri variables in the Magellanic Clouds that showed either very high ($\gtrsim$~1~M$_{\odot}$) or very low mass values ($\lesssim$0.5~M$_{\odot}$), which are in conflict with the standard single-star evolution of a post-AGB object, likely due to different evolutionary channels and perhaps revealing the effects of binarity. It may be that some of these post-AGB systems may evolve too slowly to ever become planetary nebulae \citep{vanWinckel2017}. \citet{Bodi2019} showed similar mass estimate results to those of \citet{Groenewegen_Jurkovic2017}, for a small sample of Galactic RV~Tauri variables.

It was recently discovered that some of the less luminous RV~Tauri variables are actually post-red giant branch (post-RGB) stars, a class of objects that have similar spectroscopic stellar parameters to their post-AGB counterparts but appear at lower luminosities on the H-R diagram \citep{Kamath2016, Manick2018}. These objects are thought to have evolved off the RGB instead of the AGB as a result of binary interaction.

The CBDs around post-AGB (and post-RGB) binaries are described as second-generation `scaled-up' analogs to proto-planetary disks surrounding young stellar objects \citep[YSOs;][]{Hillen2017,Vanwinckel2018}. The near-IR excess observed in the SEDs of these bright post-AGB binaries, especially the RV~Tauri variables, comes from hot dust close to the central source, where a disk-like nature of the near-IR emission has been confirmed by interferometric observations \citep{Deroo2006}. The mid- to far-IR excess then comes from cooler dust in the disk \citep{DeRuyter2006PostAGBs,Hillen2015ACHer}.

The CBDs are found to be relatively compact with sufficient angular momentum, in combination with the binary system's gravity, to settle and form the rotating disk from ejected material, especially from the huge mass loss after the AGB phase (up to $10^{-4}$~M$_\odot$~yr$^{-1}$) of the evolved primary \citep{Bujarrabal2013}. Keplerian rotation of the circumbinary material around these stars has been resolved by narrow CO line profiles, as well as the presence of large grains with a high degree of crystallinity, indicating longevity and stability, that form these disks \citep{Gielen2011} with diameters ranging between $\sim$100 and 2000~au \citep{Bujarrabal2005,Bujarrabal2013,Bujarrabal2018}.

The slower photometric phenomenon from the RVb subset of RV~Tauri systems (and in other post-AGB binaries) has been attributed to an extrinsic variable extinction. An inclined CBD can shadow the primary at certain phases of its orbit \citep[e.g.,][]{Manick2017}, blocking the light, and causing extinction and scattering along our line of sight. \citet{Vega_2017} used ultra-precise flux measurements from the \textit{Kepler} telescope and found that the decrease in pulsation amplitude of DF~Cyg (the only RV~Tauri system in \textit{Kepler}'s original field of view) perfectly tracked the decrease in flux from its RVb oscillation, showing that the long-term minima are due to the disk obscuring the pulsating primary. \citet{Kiss2017} further extended this study to all known RVbs in the Galaxy (by also adopting flux units instead of the inverse logarithmic magnitude system) to confirm that all RVbs displayed the correlation found in DF~Cyg. Furthermore, by using interferometric observations, \citet{Kluska2019} have found that the RVb sources in their sample have high inclinations, implying a very high disk scale height and allowing the disk-shadowing interpretation to be correct.

The presence of disks has also been linked to the phenomenon known as depletion, which is found in post-AGB binaries. Depletion is a systematic under-abundance of refractory elements, in the photospheres of post-AGB stars, that correlates with the condensation temperature of an element \citep{Giridhar2000ApJ}. In order to get a depleted photosphere, the stellar radiation pressure on circumstellar material separates dust grains (containing refractory elements) from the volatile-rich gas that gets reaccreted onto the stellar surface. Since dust grains experience a much larger radiation pressure, they do not get accreted. Depletion is thought to be caused by accretion of metal-poor gas from the CBD \citep{Waters1992}. Even though the presence of a disk seems to be needed, it is not a sufficient condition for depletion to occur since not all post-AGB binaries with disks are depleted \citep{Gezer2015}.

Though it is clear that binarity plays a very important role in the dynamics and evolution of post-AGB systems, the details of binary interaction processes are still not well understood \citep{Vanwinckel2018} and have been an important topic of investigation. New advances on the discovery of high-velocity outflows in post-AGB binaries have brought to light unique orbital phase-dependent variations in the H$\alpha$ profiles and have proven to be rather common in post-AGB binaries \citep{Gorlova2012_AAP_timespec_jet_bd46442,Gorlova2014,Gorlova2015,Bollen2017,Bollen2019,Bollen2020}. Studies associate this phenomenon with binary interaction and show that these high-velocity outflows (i.e. a bipolar jet) are launched by an accretion disk around the companion (i.e. circumcompanion accretion disk) that produces a P Cygni profile from the H$\alpha$ emission line, as the jet from the companion transits the bright post-AGB primary. Based on accretion models for two well-sampled post-AGB binaries, \citet{Bollen2020} concluded that the CBD is likely feeding a circumcompanion accretion disk. This agrees with the observed depletion patterns of refractory elements in post-AGB binaries \citep{Oomen2020}. \citet{Manick2019} have already shown similar H$\alpha$ variations that point to the presence of jets for two RV~Tauri systems: RV~Tau and DF~Cyg.

In the context of the recent developments in the field of post-AGB binaries, we report new observations that may provide insight into how binarity plays a role in these systems. In this paper we present the most comprehensive, multiwavelength analysis to date of a classical RVb variable: U~Monocerotis (U~Mon). The analysis includes an X-ray detection---never before reported for any RV~Tauri system---and an unprecedented photometric time series spanning 130 yr. Our results focus on consolidating several characteristics displayed in previous measurements, such as the very long time baseline photometry. We introduce recent space- and ground-based data collected for U~Mon from the XMM-Newton X-ray satellite and the Submillimeter Array (SMA), respectively. 

We give an introduction to the U~Mon system in Section~\ref{sec:umon} and describe each data set in Section~\ref{sec:data}. We present our results in Section~\ref{sec:results}, including evidence for an even longer-term secular variation in the U~Mon light curve that may recur every $\sim$60 yr and that may represent a persistent feature in the U~Mon CBD; a refined determination of the disk's inner hole, which appears to coincide with the size of the binary orbit; a new characterization of the binary companion star; and evidence for its possible interaction with the CBD's inner edge. In Section~\ref{sec:discussion} we discuss the implications of our findings and present U~Mon as a template for the discovery of X-rays in the RV~Tauri subclass of post-AGB binaries. Finally, we summarize our conclusions in Section~\ref{sec:conclusions}.

\section{The U~Monocerotis System}\label{sec:umon}

In this section we provide an observational overview of the U~Mon system and review the physical properties that have been determined from previous studies. Table~\ref{Tab:Properties} summarizes the observed and derived physical properties that we use in our analysis.

\begin{table}
\caption{Observed and Derived Physical Properties for the U~Mon System Used in Our Analysis}
\begin{center}
\begin{tabular}{rrcr}
\hline
\hline
Properties              & Value                               && Reference \\
\hline
Spectral type           & K0Ibpv                              && 1 \\
${\rm [Fe/H]}$          & $-0.8$                              && 2 \\
${\rm [C/O]}$           & $0.8$                               && 2 \\
T\textsubscript{eff}    & 5000~K                              && 2 \\
Distance                & $1111_{-102}^{+137}$~pc             && 3$^a$ \\
Radius                  & $100_{-13.2}^{+18.9}$ R$_\odot$     && 3 \\
Luminosity              & $5480_{-882}^{+1764}$ L$_\odot$     && 3 \\ 
Mass\textsubscript{Post-AGB}&$2.07_{-0.9}^{+1.4}$~M$_\odot$   && 3$^b$ \\
Pulsation formal period & $91.48$~days                        && 4 \\
RVb long-term period    & $2451$~days                         && 4 \\
\hline
\bf{Binary}             &                                     &&   \\
Inclination             & {$75^\circ$}                        && 5 \\
Orbital period          & {$2451$~days}                       && (fixed) 6 \\
$e$                     & {$0.31 \pm 0.04$}                   && 6 \\
T\textsubscript{0}      & {$2452203\pm 17$~days}              && 6 \\
$\omega$                & {$95^\circ \pm 7^\circ$}            && 6 \\      
K\textsubscript{1}      & {$13.5 \pm 0.7$~km\,s$^{-1}$}       && 6 \\
Mass function           & {$0.54 \pm 0.12$~M$_{\odot}$}       && 6 \\
Mass\textsubscript{Companion}& $2.22^{+1.0}_{-0.75}$~M$_\odot$&& 6 \\
Semimajor axis          & {$5.78_{-1.4}^{+2.7}$~au}           && 6 \\
\hline
\bf{Disk}               &                                     &&   \\
Diameter                & $\lesssim 550$~au                   && 6 \\
Dust mass & $\sim$~$4\times10^{-4}$~M$_{\odot}$               && 6 \\
Inner-disk edge radius  & $\sim$~4.5--9.0~au                   && 6 \\
\hline
\end{tabular}
\begin{tablenotes}
\small
\item References. (1) \citet{He2014}; (2) \citet{Giridhar2000ApJ}; (3) \citet{Bodi2019}; (4) \citet{Kiss2017}; (5) \citet{Oomen2018}; (6) this study. \\$^a$ Based on \textit{Gaia} measurements. However, see note in text (Section~\ref{sec:binary}) regarding the \textit{Gaia} parallax accuracy and uncertainty. \\ $^b$Average of the values from the \citet{Groenewegen_Jurkovic2017} period--luminosity--mass relationships adopted in \citet{Bodi2019}.
\end{tablenotes}
\label{Tab:Properties}
\end{center}
\end{table}

\subsection{U~Mon as an RV~Tauri Variable of RVb Type}

U~Mon, a yellow supergiant variable \citep[K0Ibpv;][]{He2014}, is located at a distance of $1111_{-102}^{+137}$~pc \citep{Bodi2019}. It is the second-brightest RV~Tauri variable (after R~Sct) with a magnitude range of 5.45--7.67 in V \citep{Watson2006}, mean color index $B - V = 1.05$~mag at RVb brightness maximum, $1.11$~mag at RVb minimum \citep{Pollard1996}, and a metallicity of [Fe/H] $= -0.8$ \citep{Giridhar2000ApJ}. Its formal pulsation period (deep + shallow minima cycle) is 91.48~days and its fundamental period is 45.74~days. U~Mon also exhibits a long-term periodic modulation in mean brightness (RVb phenomenon) with a period of $\sim$2451~days \citep{Kiss2017}, where the large-amplitude difference in mean brightness is $\sim$3 mag.

\subsection{U~Mon as a Binary Star System}\label{sec:binary}

With BVRI photometry and high-resolution spectra, \citet{Pollard1995} concluded that U~Mon is an eccentric binary (e = 0.43) exhibiting a radial velocity amplitude of 30~km~s$^{-1}$ (full amplitude) and an orbital period of $\sim$2597~days, which is similar to its photometric RVb period, as found in other RVb systems \citep{Manick2017}.

\citet{Oomen2018} recently updated the orbital properties of the U~Mon binary system using data from the HERMES spectrograph on the 1.2 m Mercator telescope. They found an orbital period of $2549 \pm 143$~days. By assuming a typical post-AGB mass for the primary of 0.6~M$_\odot$ and an inclination of $75^{\circ}$, they estimated a projected semi-major axis of $3.38 \pm 0.31$~au, a mass function of $0.79 \pm 0.18$~M$_\odot$, and hence a minimum mass of 1.64~M$_\odot$ for the companion. However, most recently \citet{Bodi2019} used \textit{Gaia} \citep[DR2;][]{Gaia2018} data and two different period--luminosity--mass--temperature--metallicity relations derived by \citet{Groenewegen_Jurkovic2017} (based on hydrodynamical atmosphere modeling of fundamental-mode pulsators from \citet{Bono2000,Marconi2015}), to empirically infer the mass of the post-AGB star in U~Mon to be between 2.00 and 2.13~M$_\odot$. For our analysis we adopt the mean of these last two values (Section~\ref{ssec:orbital_properties}).

We note that the \textit{Gaia} distance is likely biased by the orbital movement of the binary. The renormalized unit weight error (RUWE) for U~Mon (RUWE = 2.4) indicates a poor single-star solution. High RUWE can be caused by a variety of factors, including resolved components in U~Mon \citep[see][]{2020MNRAS.496.1922B}. The astrometric excess noise ($\epsilon_{Gaia}$) for U~Mon in the DR2 catalog is 0.40~mas. This parameter is the excess uncertainty that must be added in quadrature to the formal uncertainties to obtain a statistically acceptable astrometric solution in the DR2 pipeline \citep{Lindegren2012}. \citet{Gandhi2020} considered using  $\epsilon_{Gaia}$ as a proxy for the expected astrometric orbital wobble ($\omega$) to help identify potential X-ray binaries. Given that the \textit{Gaia} DR2 observations\footnote{Section~1.3.1 Time coverage: \url{https://gea.esac.esa.int/archive/documentation/GDR2/}} occurred over only $\sim$25$\%$ of the orbit of the U~Mon system, the maximum expected orbital wobble for our orbit solution for U~Mon at a distance of 1.1~kpc is $\sim0.3$~mas, which is comparable to the reported $\epsilon_{Gaia}$ value, suggesting that the excess uncertainty could be mostly due to orbital motion. Indeed, the newly updated \textit{Gaia} Early Data Release 3 parallax\footnote{\textit{Gaia} EDR3 was released on 2020 December 01.} of 1.28$\pm$0.12~mas differs from the \textit{Gaia} DR2 parallax by 0.36~mas, again very similar to $\epsilon_{Gaia}$. Therefore, if we adopt $\epsilon_{Gaia}$ as the uncertainty in the parallax measurement, then the distance to U~Mon ranges between $\sim$770~pc and $\sim$2~kpc. However, using the same period-luminosity-mass relations by \citet{Groenewegen_Jurkovic2017} as were used originally by \citet{Bodi2019}, we find that the stellar and orbital quantities in Table~\ref{Tab:Properties} do not change significantly relative to the already-quoted uncertainties. Therefore, we adopt the nominal distance of 1.1~kpc as in \citet{Bodi2019}.

\subsection{The U~Mon Circumbinary Disk}

As with many other RV~Tauri variables in the literature, the U~Mon binary star system is surrounded by a CBD. \citet{Bujarrabal2013} estimated the size of the U~Mon CBD to be $\sim$300~au using observations at the IRAM 30 m telescope in 2012--2013. Though they did not detect the Keplerian rotation of the CBD, they derived an upper limit for the molecular gas mass ($<$9~$\times$~$10^{-4}$ M$_\odot$) from their \mbox{$^{13}$CO} \mbox{$J$=1$-$0} data.

In addition, \citet{Kluska2019} were able to reproduce their recent Very Long Baseline Interferometer (VLTI)/PIONIER near-IR (H-band) observations for U~Mon by using their most complex model that includes a binary and inner ring. They suggested an an inner-disk diameter of $\sim$5.9~AU.

\subsection{Magnetic Activity in U~Mon}

\citet{sabin2015} were the first to find magnetic fields at the surface of U~Mon. The Stokes~Q and U profiles were observed at RVb phase~$\sim$0.63 and the Stokes~V profile at phase~$\sim$0.84 in Figure \ref{fig:lc_min_var}, respectively. They analyzed high-resolution spectropolarimetric ESPaDOnS (CFHT) data and found a clear Zeeman signature in the Stokes~V profile. They measured a longitudinal magnetic field (i.e., in the line of sight) of $10.2\pm1.7$~G in the photosphere of U~Mon. Although the sample is poor, this is to date the strongest surface field directly detected for a post-AGB star. The Stokes~Q and U profiles indicated the presence of shocks, and the authors suggested the possible amplification of the magnetic field due to the atmosphere dynamics. 

\section{Data}\label{sec:data}

\subsection{Radial Velocity Observations}\label{ssec:rv}

\begin{figure}[!ht]
    \centering
    \includegraphics[width=\linewidth,trim=100 70 70 80,clip]{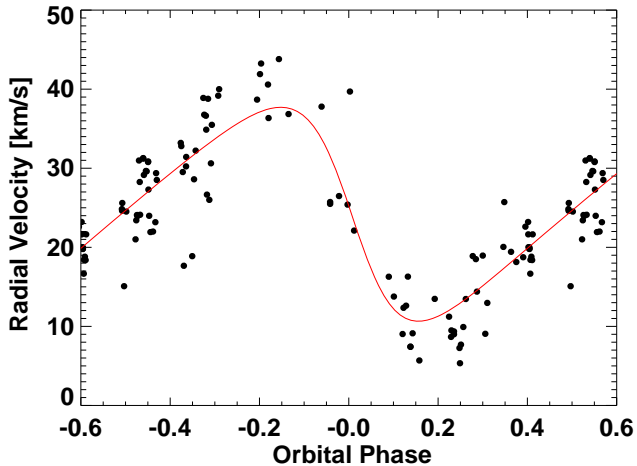}
    \caption{Single-lined orbit solution for U~Mon as fit to the radial velocity observations of \citet{Oomen2018}.}
    \label{fig:orbit}
\end{figure}

We adopted radial velocity observations of U~Mon, corrected for the effects of surface pulsations, reported by \citet[][available on VizieR]{Oomen2018}. Most photometric RVb periods of RV~Tauri variables are similar to the orbital periods of their binaries \citep{Manick2017}. To calculate a new orbital solution for the binary in U~Mon, we fixed the orbital period at 2451~days \cite[U~Mon's photometric RVb period;][] {Bodi2019}, and we refit the data with a single-lined spectroscopic, Keplerian binary orbit model using {\tt PHOEBE} \citep{Prsa2016}. The resulting fit is shown in Figure~\ref{fig:orbit}, and the fit parameters are summarized in Table~\ref{Tab:Properties}. 
We discuss our findings in Section~\ref{ssec:orbital_properties}.

\subsection{Light-curve Observations}

\begin{figure*}[!ht]
    \centering
    \includegraphics[width=\linewidth]{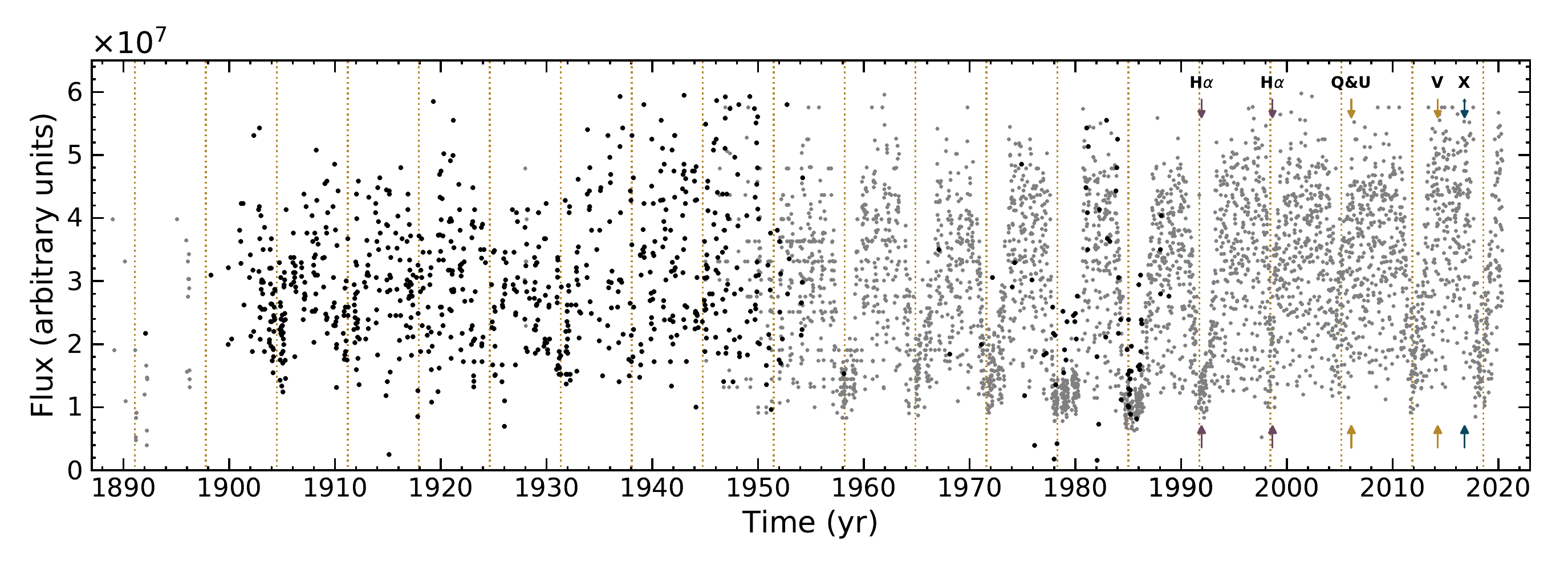}
    \caption{Final combined AAVSO (gray points) and DASCH (black points) light curve of U~Mon, binned by 5 days. Both data sets overlap in time between 1945 and 1954, and during two long-term RVb cycles between 1974 and 1988. The scatter is due to the short-term pulsation variability. The times of apastron passage are represented by golden vertical dotted lines; the apastron times match with the RVb minima throughout the entire light curve (see Section~\ref{ssec:orbital_properties}). The arrows represent the observation times of the enhanced H$\alpha$ (purple); the Stokes Q, U, and V profiles (gold); and X-rays (teal). The light curve in this plot is available as Data behind the Figure (DbF).}
    \label{fig:umon_lc}
\end{figure*}

We use two sources of long-term photometric monitoring of U~Mon in order to explore secular changes in the U~Mon light curve on timescales of decades or longer. The full light-curve data set is represented in Figure~\ref{fig:umon_lc}.

\begin{figure}[!ht]
    \centering
    \includegraphics[width=0.5\textwidth]{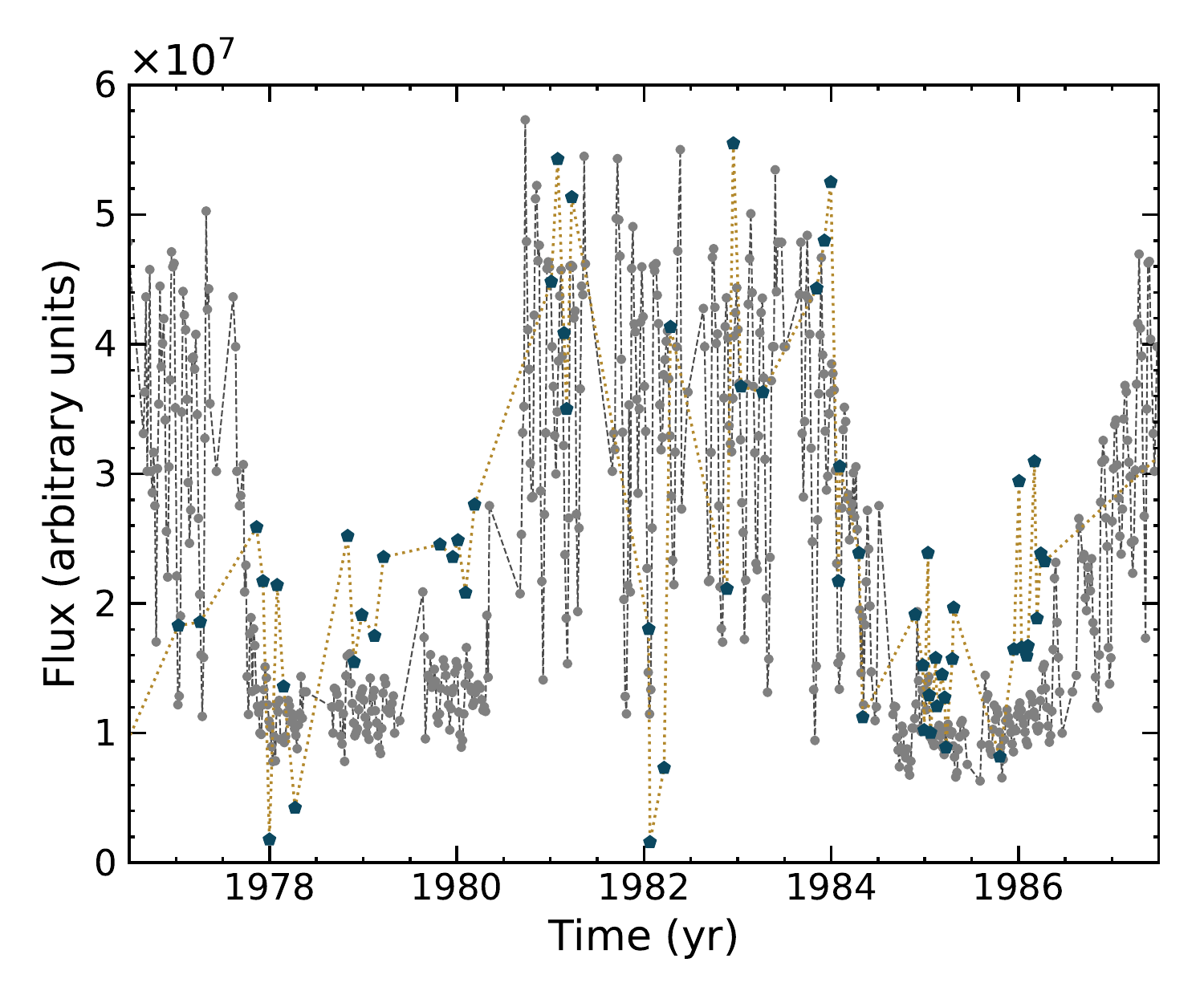}
    \caption{Two long-term RVb minima where the DASCH (teal pentagons) and AAVSO data (gray points) overlap the most in time. The grey dashed line on the AAVSO points (yellow dotted line for DASCH), though affected by the 5-day binning of the data, mainly highlights the pulsation variation of U~Mon. Pulsation amplitudes during the RVb maxima may extend to as low as the mean magnitude at RVb minima, while the pulsations during RVb minima are always smaller in amplitude, showing the effect of disk obscuration of the pulsating post-AGB.}
    \label{fig:umon_zoom_lc}
\end{figure}

\subsubsection{{\normalfont AAVSO}}
The American Association of Variable Star Observers (AAVSO) is a global network of amateur and professional astronomers dedicated to monitoring variable stars. The earliest observation of U~Mon in the AAVSO archive was made by Ernest E. Markwick on 1888 December 25 (JD 2,410,997.0), and after a few observations there followed a 49 yr gap between 1896 and 1945. More regular monitoring began in the mid-1940s and continues to the present day. We downloaded all (including ``discrepant"-flagged) Mag$_{V}$ data from the AAVSO database up to 2020 May 26. We chose to keep discrepant data since we are not focused here on the pulsation variability. We also wanted to make sure we had the data between 1888 and 1896 (which are all marked discrepant) because they capture a long-term RVb cycle; these data are nonetheless useful for our purposes and fill a gap in time that is otherwise unavailable. We only excluded 36 Mag$_{V}$ data points tagged as upper limits.

\subsubsection{{\normalfont DASCH}}
The Digital Access to a Sky Century at Harvard (DASCH) survey is an ongoing effort to digitize about 0.5 million photographic plates covering the northern and southern sky from 1880 to 1985 \citep{Grindlay2009}. 

Data for U~Mon were released in the DASCH Data Release 6 (DR6). 
We downloaded the light curve of U~Mon from the DASCH Light Curve Access pipeline website\footnote{\url{http://dasch.rc.fas.harvard.edu/lightcurve.php}} using the default search radius of $5\arcsec$. Since the majority of the Harvard plates are close to Johnson B, we chose the data from the APASS B photometric calibration catalog, which yields the most accurate photometry \citep{Tang2013}.

The DASCH (APASS B-band catalog) light curve of U~Mon (ID: T540046991) contained 3824 magnitude data points from approximately 3436 plates at the time of download. The light curve has a mean magnitude of 8.77 in B and a baseline going back to 1888 January 25 (JD 2410661.7) with occasional gaps, the largest one being the ``Menzel'' gap in the 1950s and 1960s, when the plate-making operation was halted by the Harvard Observatory director at the time owing to financial concerns. We excluded DASCH data points that had estimated errors of the locally corrected magnitude measurement ({\verb magcal_local_rms }) values greater than 0.6 mag (which included {\verb magcal_local_rms } values set to 99.0; these are tagged magnitudes from images dimmer than the limiting value of the image). We also excluded magnitude-dependent corrected magnitude ({\verb magcal_magdep }) values brighter than 2.0 and dimmer than 11.0 (which were well away from the mean magnitude of the overall light curve).

\subsubsection{Combining {\normalfont AAVSO} and {\normalfont DASCH} Light-curve Data}

To analyze the AAVSO and DASCH data sets together, we converted the light curves to flux units using arbitrary zero-points (ZPs). For AAVSO we adopted ZP~=~25 \citep{Kiss2017}. Because the DASCH effective bandpass is different from the AAVSO Mag$_V$ bandpass, we experimented with ZP values for DASCH to empirically determine a scaling factor and ensure a proper match of the AAVSO and DASCH data during two RVb cycle amplitudes (between 1974 October and 1988 August), where the AAVSO and DASCH data sets overlap to the greatest degree in time. We adopted ZP~=~27.5 for DASCH as the best fit but found that we also had to scale the DASCH fluxes by a factor of 1.21 and then subtract $1.1\times 10^{7}$ flux units to match the two overlapping RVb cycles (see Figure~\ref{fig:umon_zoom_lc}). After we converted to arbitrary flux units, we made another cut to exclude data above $\sim$ $6 \times 10^{7}$ flux units and $\leq$0.

The final combined AAVSO+DASCH light curve is shown in Figure~\ref{fig:umon_lc}. It is an impressive data set, spanning the period from $\sim$1890 to $\sim$2020. The light curve is binned by 5~days to average out errors of individual observations and to reduce phase smearing due to the binning \citep{Kiss2017}. What seems to be leftover scatter is actually due to pulsations of different amplitudes, as well as large gaps in the data (see, e.g., Figure~\ref{fig:umon_zoom_lc}). Several features are present in the secular changes of the light curve on timescales of decades, and we discuss this in Section~\ref{ssec:secular_variations}.

\subsection{{\normalfont SMA} Observations}

The SMA is an interferometer, composed of eight 6 m dishes, that observed U~Mon on three occasions. On 2018 February 1, we observed the source in the subcompact array configuration with projected baselines of 7--49~m and covered two frequency ranges at 223.6--231.6~GHz and 239.6--247.5~GHz. On 2018 October 2, observations were made at higher frequencies, from 328.8 to 360.8~GHz, and at longer baselines of 6--70~m. 
On 2019 March 15, we finally observed U~Mon in the very extended SMA configuration (VEX) with baselines from 32 to 514~m. The covered frequency ranges were 209.1--212.8~GHz, 214.9--216.7~GHz, 225.1--226.9~GHz, 229.2--238.8~GHz, 332.8--340.8~GHz, and 348.8--356.8~GHz. The spectral coverage in the VEX observations is smaller than on the earlier dates owing to a nonfunctional quadrant of the SMA correlator SWARM \citep{SWARM} at that time. All eight SMA antennas were used for our first two observing runs except for observations in VEX at low frequencies for which only six antennas had properly functioning receivers. 

Because the target was observed mainly as a filler project,\footnote{\url{http://sma1.sma.hawaii.edu/call_filler.html}} the final {\it uv} coverage is suboptimal, in particular, very inhomogeneous at higher frequencies. The complex antenna gains were calibrated in all observing runs by observations of quasars J0730--116 and J0725--009. The bandpass was calibrated using long integrations of quasars 3C 279 and 3C 84. An absolute flux scale was established by observations of Uranus (on February~1), Neptune (on October~2), and Callisto (in 2019). The calibration was performed in {\tt MIR} software using standard procedures\footnote{\url{http://www.cfa.harvard.edu/~cqi/mircook.html}}. Further data processing, including imaging, was performed in {\tt CASA} \citep{CASA}. 

\begin{figure*}[!ht]
    \centering
    \includegraphics[width=\linewidth]{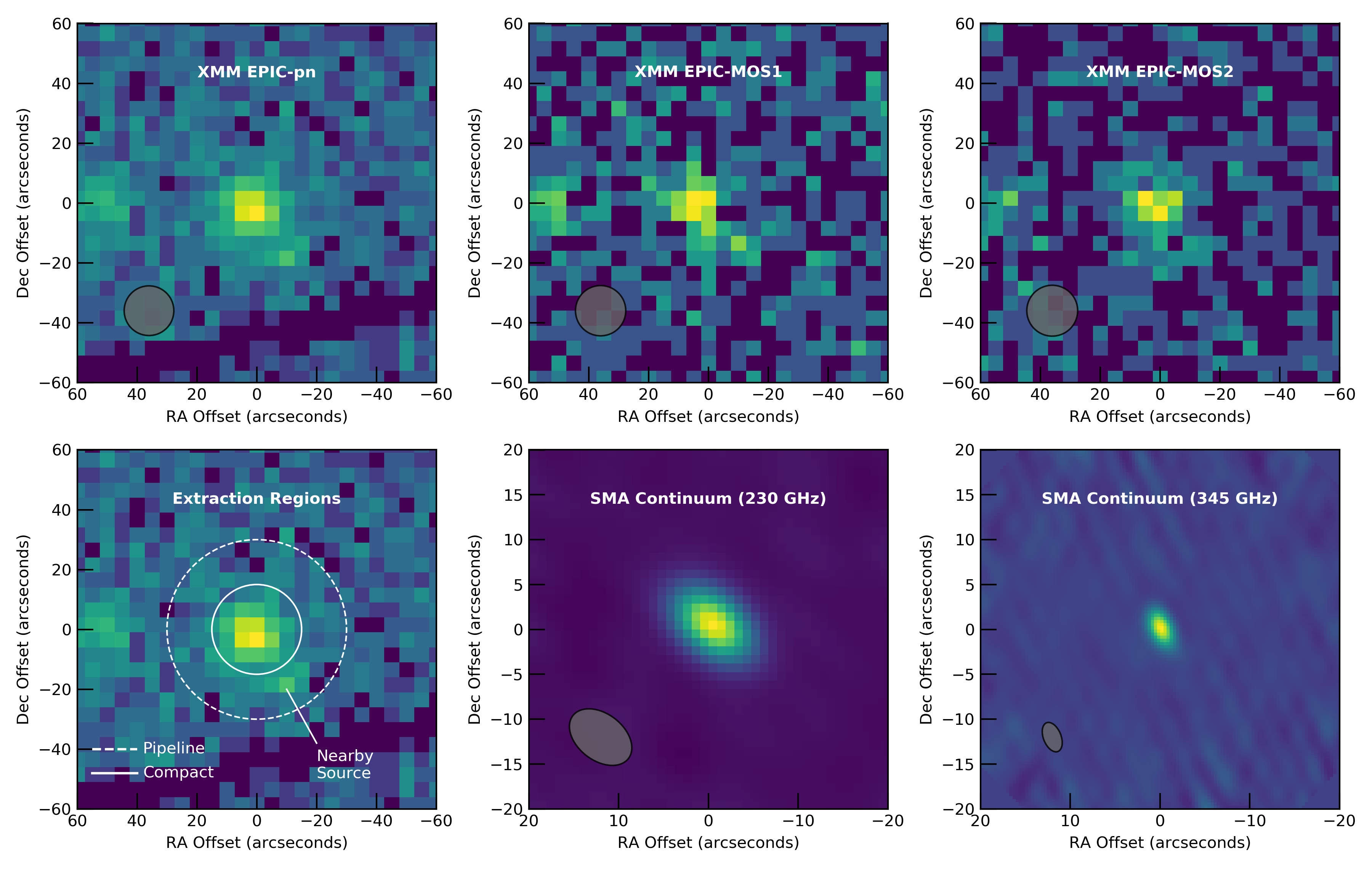}
    \caption{Top panels: XMM EPIC-pn, EPIC-MOS1, and EPIC-MOS2 images with a clear detection of X-rays for U~Mon. Bottom left panel: The EPIC-pn image annotated with the pipeline extraction vs. our compact extraction region excluding the nearby source near U~Mon. Bottom middle and right panels: U~Mon's SMA continuum images at 230 GHz (1.3~mm) and 345 GHz (870 $\mu$m), respectively. The respective beams are included in the lower left corners in dark gray.}
    \label{fig:umon_images}
\end{figure*}

Calibrated visibilities measured in the different array configurations were combined within the two atmospheric windows near 345~GHz ($\sim$0.8 mm) and 230~GHz (1.3~mm) and their weights rescaled to the actual noise levels. This resulted in continuum sensitivities of $\sigma_{345}=4.0$~mJy~beam$^{-1}$ and $\sigma_{230}=0.89$~mJy~beam$^{-1}$ at beam sizes of  4\farcs0~$\times$~2\farcs4 and 0\farcs9~$\times$~0\farcs6, respectively. These beam sizes correspond to natural weighting of visibilities. In both bands the continuum source was readily detected at flux levels listed in Table~\ref{Tab:SED}; see also Figure~\ref{fig:umon_images}. No spectral lines were detected with an rms noise level of 39.3~mJy~beam$^{-1}$ near the frequency of the CO $J$=2--1 line and at a 3.9~km~s$^{-1}$ spectral binning, which is consistent with the nondetection reported by \citet{Bujarrabal2013}. We discuss U~Mon's submillimeter emission in Section~\ref{ssec:sma_umon}.

\subsection{{\normalfont XMM}-Newton Observations}\label{ssec:data_xray}

U Mon was observed by the X-ray Multiple Mirror (XMM-Newton) observatory on 2016 October 23 for 58.3 ks (RVb phase $\sim$0.23 in Figure \ref{fig:lc_min_var}). The XMM-Newton observations include 15 imaging exposures: three X-ray images on the EPIC-pn (55.3 ks), EPIC-MOS1 (57.0 ks), and EPIC-MOS2 (56.9 ks) detectors, and 12 exposures (each $\sim$2.2--4.4 ks) with the Optical Monitor (OM) using the UVW1, UVM2, and UVW2 filters. Representative images are shown in Figure~\ref{fig:umon_images}. U Mon is detected in all imaging exposures. Additionally, there are two grating dispersed exposures in X-ray emission on the RGS1/2 (57.2 ks), but the dispersed spectrum is not detected. The large source region used to generate the pipeline products leads to confusion and blending with a nearby source (see bottom left panel in Figure~\ref{fig:umon_images}). We re-extracted the X-ray spectra and light curves using a smaller source region with a radius of $15^{\prime\prime}$ and a source-free background region near the source. The extracted spectral products were corrected for the reduced extraction region by accounting for the encircled energy fraction in the response files. 

\subsection{Spectral Energy Distribution Data}\label{sec:SED}

To construct U~Mon's SED, we downloaded available archival photometric data from VizieR. We did not take into account the phase of observations (i.e. either at maximum/minimum pulsation or long-term RVb brightness). 
We included Herschel-PACS/SPIRE measurements (downloaded from the ESA Herschel Science Archive\footnote{\url{http://archives.esac.esa.int/hsa/whsa/}}; PI: C. Gielen), the $850~\micron$ flux value reported by \citet{Deruyter2005RVTau} taken with the Submillimetre Common-User Bolometer Array (SCUBA) at the James Clerk Maxwell Telescope (JCMT), and finally millimeter fluxes from \cite{Sahai2011} taken at the Combined Array for Research in Millimeter-wave Astronomy (CARMA).

\begin{figure*}[!ht]
    \centering
    \includegraphics[width=\linewidth,trim=10 15 10 10,clip]{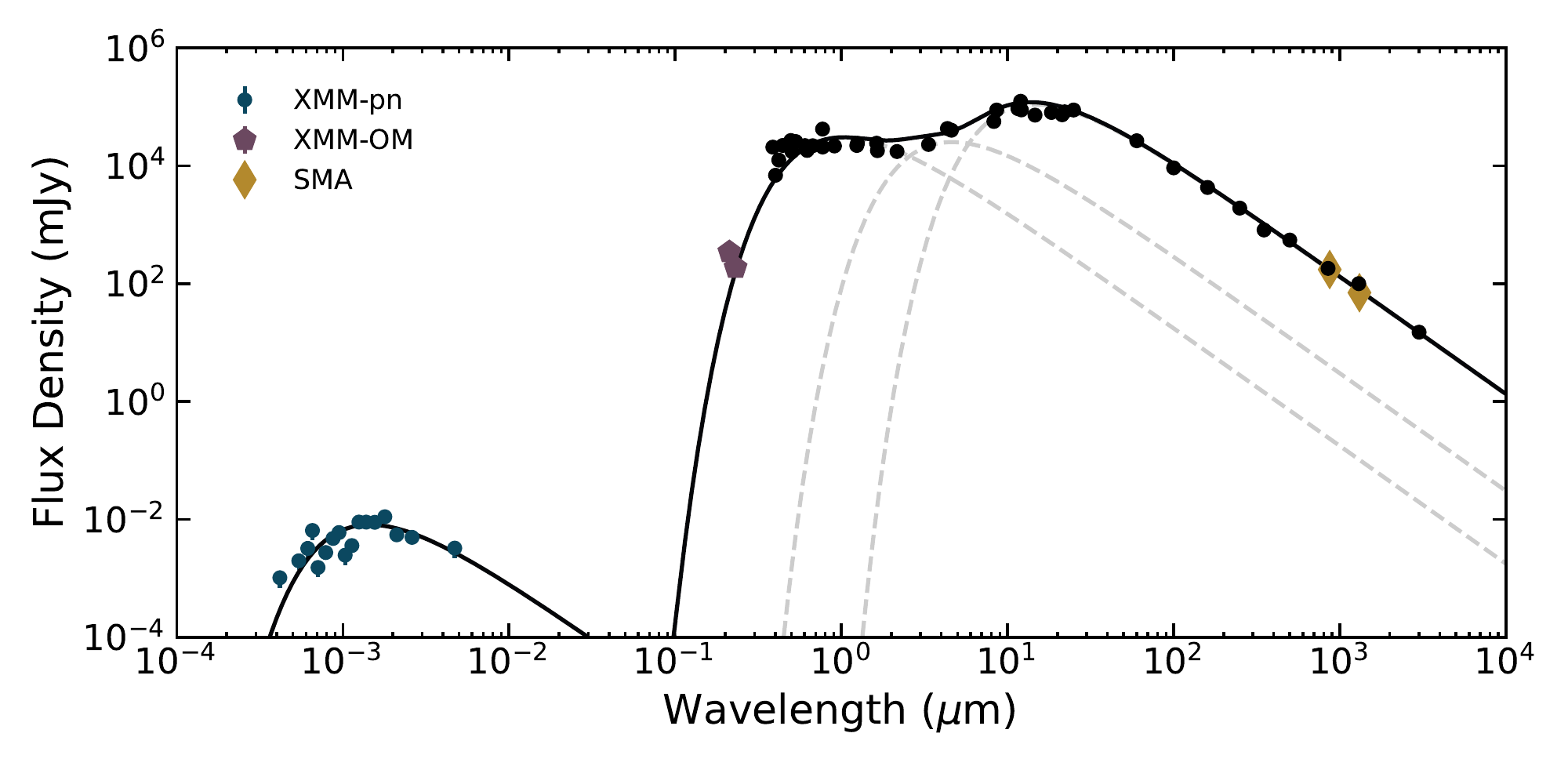}
    \caption{\small SED for U Mon. Representative archival data (black points) were collected from VizieR and the Herschel archives, as well as the 850~$\mu$m measurement from \citet{Deruyter2005RVTau} and the 1.3 and 3~mm measurements from \citet{Sahai2011}. We present the new XMM-pn spectrum in teal points, the XMM-OM data in pentagons, and the SMA data in diamonds. 
    The scatter in the archival data is probably due to intrinsic (pulsation or RVb) variability of U~Mon at the time of observation. 
    Note: error bars are smaller than the symbols.}
    \label{fig:UMon_SED}
\end{figure*}

\begin{table*}[!ht]
\caption{New Flux Measurements for U~Mon}
\begin{center}
\begin{tabular}{cccc}
\hline \noalign {\smallskip}
Wavelength ($\micron$) & Flux (mJy) & Date & System \\
\hline \noalign {\smallskip}
0.212 & 23.27 $\pm$ 0.038 & 2016 Oct & XMM-OM:UVM2\\
0.231 & 16.30 $\pm$ 0.030 & 2016 Oct & XMM-OM:UVW2\\
869.4 &173.3  $\pm$ 6.5   & 2018 Feb, Oct; 2019 Mar\footnote{\label{tablenote:a} Combined observations.} & SMA\\
1313  & 70.2 $\pm$ 8.5   & 2018 Feb, Oct; 2019 Mar$^{\ref{tablenote:a}}$ & SMA\\
\hline
\end{tabular}
\label{Tab:SED}
\end{center}
\end{table*}

The new observations we contribute in this paper are submillimeter and UV/X-ray from the SMA and XMM-Newton, respectively. Our SMA $870~\micron$ and 1.3~mm values are similar to those reported by \citet{Deruyter2005RVTau} and \cite{Sahai2011}, respectively. The XMM-OM magnitudes for U~Mon were converted to flux density (Jy) using the ZPs from \citet{Mason2001}. To represent the XMM EPIC-pn spectrum, we unfolded the instrumental response from the spectral data assuming the best-fit model determined using the X-ray Spectral Fitting Package \citep[{\tt XSPEC}\footnote{\url{http://heasarc.gsfc.nasa.gov/xanadu/xspec/}};][]{Arnaud1996}, dereddened the spectral data (see Section~\ref{ssec:results_SED}), and then converted from photon flux densities to Jy. 

Figure~\ref{fig:UMon_SED} shows the SED for U~Mon, and Table~\ref{Tab:SED} lists our new flux measurements. In Section~\ref{ssec:results_SED} we discuss the blackbody models we used on the SED.

\section{Results}\label{sec:results}

\subsection{Secular Variations of the U~Mon Light Curve over the Past Century}\label{ssec:secular_variations}

The long-term RVb cycles of the U~Mon light curve are more salient in the AAVSO data, and we can estimate the duration of the maxima more easily than the minima. Overall, the RVb phenomenon in U~Mon changes significantly from cycle to cycle. The RVb maxima range between $\sim$3.3 and 4.8~yr, whereas the RVb minima differ even more, from several months to the longest minimum lasting $\sim$~2.5~yr (e.g., between $\sim$1977.8 and 1980.5; see Figure \ref{fig:umon_zoom_lc}). 

\begin{figure*}[!ht]
    \centering
    \includegraphics[width=\linewidth]{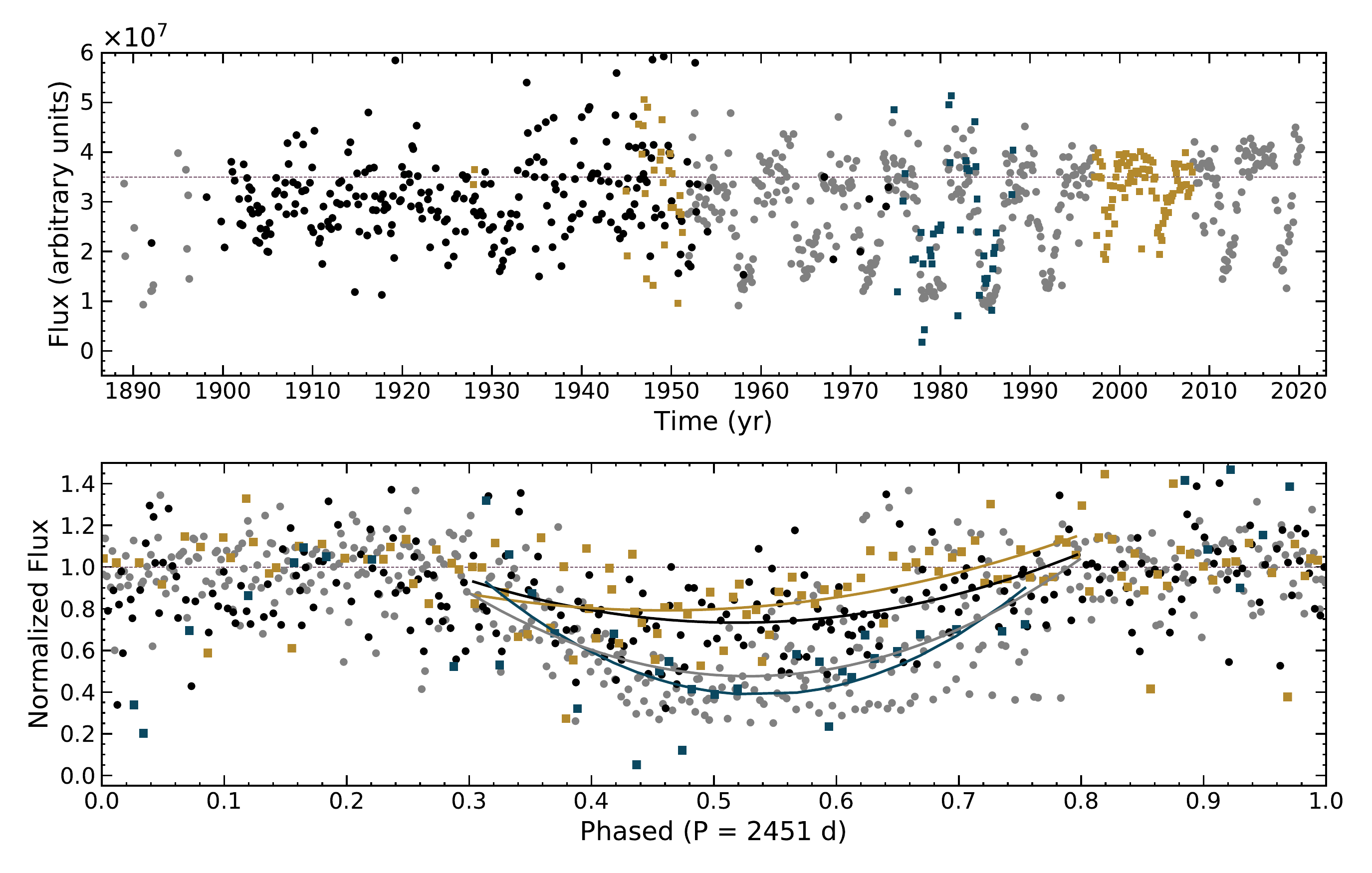}
    \caption{\small Analysis of secular variations in the U~Mon light-curve data. Top: light curve binned by the fundamental period 45.74~days. The black points are DASCH data, the gray points are AAVSO data, the DASCH data overlapping two distinct RVb cycles are shown as teal squares, and the golden squares define AAVSO data sporadically overlapping DASCH data before 1951 and data corresponding to what might be times of partially obscured minima ($\sim$1997--2008). The dashed line is the mean flux at RVb maximum ($3.5\times 10^{7}$ flux units). Bottom: light curve phase-folded on the 2451-day RVb period of the system with flux normalized at unity at the base level (dashed line is the same as the top panel). See Section~\ref{ssec:secular_variations} for a discussion of the curves.}
    \label{fig:lc_min_var}
\end{figure*}

In Figure~\ref{fig:lc_min_var} we explore longer-term secular changes in the light curve, enabled by unprecedented coverage spanning $\sim$130~yr. We binned both data sets by U~Mon's fundamental pulsation period (45.74~days) to focus on the RVb behavior. We plot the DASCH data in black, the AAVSO data in gray, and the DASCH data that distinctly overlap the two RVb cycles during 1975--1988 as teal squares. We also define golden squares that are AAVSO data sporadically overlapping with DASCH data before 1951, and AAVSO data corresponding to what might be times of partially obscured RVb minima ($\sim$1997--2008). The top panel is the binned light curve in time, the bottom panel is the binned light curve phased by the RVb period of 2451~days, and the flux is normalized to be unity at the base level. We fit parabolas to the data at phases of $0.5\pm 0.3$, to get a cleaner visual sense of how the RVb minima compare for the various subsets.

We find that the AAVSO data (gray points) and the DASCH overlap data (teal squares) have nearly identical RVb minima. Similarly, we see that the older DASCH data (black points) and the golden-square AAVSO data have nearly identical RVb minima, and the latter minima are about a factor of 2 less deep. In other words, it appears from the historical record that the RVb variations---which have come to be associated with a highly inclined disk shadowing the primary at certain orbital phases---have largely disappeared at least twice in the past, with a timescale of $\sim$75~yr, from the middle of the black points to the latest golden squares in Figure~\ref{fig:lc_min_var}. 

\begin{figure*}[!ht]
    \centering
    \includegraphics[width=\linewidth]{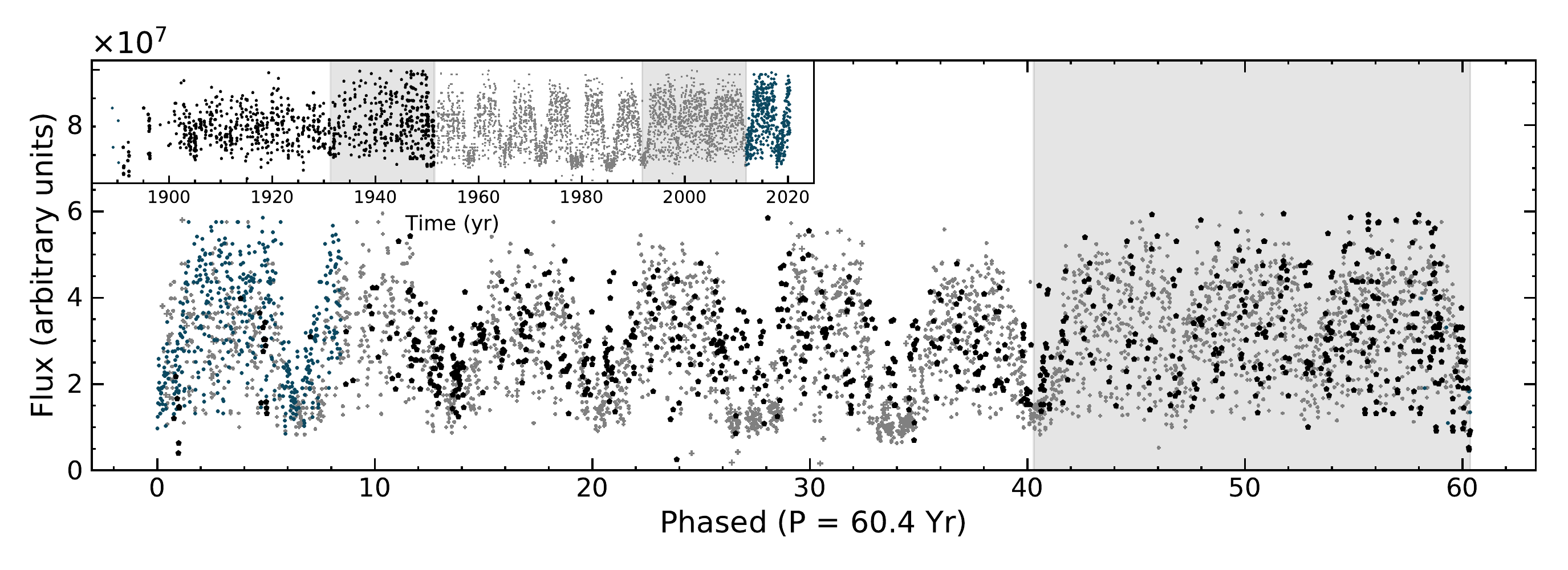}
    \caption{\small Light curve of U~Mon phased at 60.4 yr (nine RVb cycles). The inset shows the full light curve divided in color distinguishing the two 60.4 yr cycles in the data (black points represent the first 60.4 yr cycle; gray points, second cycle). The start of a third cycle is shown as teal points. The RVb cycles highlighted in light grey include long ($\lesssim$4.5~yr) RVb maxima and short RVb minima (phase~$\sim$47~yr and $\sim$53~yr) that have durations of only several months based on the AAVSO data.}
    \label{fig:lc_phased_60yr}
\end{figure*}

Another way to examine this is shown in Figure~\ref{fig:lc_phased_60yr}, in which the disappearance of the RVb variation, due to the large scatter in the data points, is more obvious for two RVb minima in 1938 and 1944 and then again in 1998 and 2005, with a timescale between them of $\sim$60.4~yr, which we note is nine times the RVb long period. To be clear, the RVb minima are short during these cycles ($\sim$several months, as opposed to years), and they ``disappear'' relative to the large scatter/pulsation variations from the (particularly longer) maximum state at either side of RVb minima. We further discuss the interpretation of this phenomenon in Section~\ref{ssec:60yr_trend}.

\subsection{Orbital Properties of the U~Mon Binary Star System} \label{ssec:orbital_properties}

From the orbit fit in Section~\ref{ssec:rv}, we obtained orbital parameters (listed in Table \ref{Tab:Properties}) that are within the errors of the values consistent with \citet{Oomen2018}, including a new periastron time of T$_{0} = 2452203 \pm 17$~days. Adding half (1225~days) of the fixed orbital period to the periastron time gives an apastron at JD $= 2453428$~days. Multiples of the orbital period with the apastron time give all apastron events (denoted as golden vertical dashed lines in Figure~\ref{fig:umon_lc}) that roughly align with the RVb minima throughout U~Mon's entire DASCH+AAVSO light curve. Mass estimates for U~Mon range from a fiducial post-AGB mass average of 0.6~M$_{\odot}$ \citep{Gezer2015,Manick2017,Weidemann1990} to most recent values of $2.00$ and 2.13~M$_{\odot}$ reported by \citet{Bodi2019}. To estimate a value for the companion's mass (M$_{2}$), we used the average value from \citet{Bodi2019} for the mass of the post-AGB (M$_{1}$ = 2.07~M$_{\odot}$) in Equation \ref{Eq:massfct} (the binary mass function for an eccentric orbit) and solved for M$_{2}$: 
\begin{equation}
f(m) = \frac{P K_1^3}{2\pi G}(1-e^2)^{\frac{3}{2}} = \frac{M_2^3}{(M_1+M_2)^2} \sin^3i.
\label{Eq:massfct}
\end{equation}

Using the recalculated radial velocity curve, and keeping the orbital inclination maximum limit of $75^\circ$ \citep{Oomen2018}, we find the mass for the companion to be $2.2_{-0.75}^{+1.0}$~M$_\odot$; this mass range corresponds to an F-type or A-type star on the main sequence. Then, by using Kepler's third law, we derived a value of $5.78_{-1.4}^{+2.7}$~au for the semi-major axis. These parameters are reported in Table~\ref{Tab:Properties}. 

\subsection{Spectral Energy Distribution}\label{ssec:results_SED}

We dereddened the flux data between 0.2 and 3.3~$\mu$m using functions contained in the {\tt Astropy}-affiliated package {\tt dust\_extinction}, with the F19 Milky Way R(V) dependent extinction \citep{Fitzpatrick2019} using the standard value Rv = 3.1 and total extinction: E\textsubscript{B-V}~=~0.3 \citep{Deruyter2005RVTau}.

We made use of the {\tt astropy.modeling.blackbody} routine \citep{Astropy2013} to depict scaled blackbody components on the SED (these are not fitted). We include two blackbody components with temperatures of 5000 and 1100~K to represent the stellar atmosphere \citep{Bodi2019} and the inner-disk edge (see Section~\ref{ssec:indisk_properties}), respectively. We also include a modified blackbody \citep[i.e., ``graybody'';][]{Casey2012} component with a mean temperature of 350~K to represent the dust in the extended disk. Finally, we added a 3.5~MK thermal blackbody for the X-ray spectrum.\footnote{The blackbody model is only shown for display purposes, the unfolded spectrum has a strong dependence on the best-fit 10~MK plasma model, and a blackbody model does not provide a suitable fit to the observed X-ray spectrum.} Figure \ref{fig:UMon_SED} shows the entire SED model as a black solid line, with the individual components depicted as light-gray dashed lines. 

We integrated the 5000~K blackbody model (between~$\sim$0.2 and 2~$\mu$m), which gave a (dereddened) bolometric flux of F\textsubscript{bol} $\approx 1.44 \times 10^{-7}$ erg~s$^{-1}$~cm$^{-2}$, corresponding to L$_{\rm \star}$ $\sim$5556~L$_{\odot}$ and yielding a physical radius of~R$_{\rm \star}$ $\sim$100~R$_{\odot}$.

We tested how an additional blackbody component with the properties of the putative A-type companion star (9800 K and 55 L$_{\odot}$) would influence our SED model and found that such a companion would remain undetectable in the SED (even at RVb minimum) owing to the brightness of the post-AGB component (see Section~\ref{ssec:discussion_xrays}).

Assuming that the dust is optically thin at $\lambda =$ 870 $\mu$m, we used the following expression \citep[e.g.][]{Hildebrand1983} to estimate the mass of the dust in the disk:

\begin{equation}
M_{d} = \frac{F\textsubscript{$\lambda$} D^{2}}{\kappa_{\lambda}B\textsubscript{$\lambda$}(T_{d})},
\end{equation}
where B\textsubscript{$\lambda$}(T\textsubscript{d}) is the blackbody intensity at the Rayleigh-Jeans limit in the form of $\frac{2k_{SB}T}{\lambda^2}$. Using a mean dust temperature T\textsubscript{d} = 350 K, our observed SMA flux F\textsubscript{870} = 173.3 mJy, D = 1.1 kpc \citep{Bodi2019}, and a dust opacity of $\kappa_{\lambda} = 2.0$ cm$^{2}$g$^{-1}$ at 870 $\mu$m \citep{Ladjal2010}, we estimate a total dust mass of M\textsubscript{d}~$\sim 4 \times 10^{-4}$~M$_{\odot}$. For a typically assumed gas-to-dust ratio of 200 \citep[e.g.][]{Groenewegen2007,Groenewegen2009} the total dust mass suggests a gas mass of $\sim$8~$\times 10^{-2}$~M$_{\odot}$; however, this gas mass is much larger than the constraint of the molecular mass estimated by \citet{Bujarrabal2013}. The lack of molecular gas may be related to high-energy radiation that has been dissociating molecules present in the system that would support the larger distance to U~Mon. Since the primary star is not hot enough, the presence of a hotter companion or interstellar UV field could photodissociate molecules in the inner and outer disk regions \citep{Bujarrabal2013}.

\subsection{Properties of the Circumbinary Disk's Inner Edge}\label{ssec:indisk_properties}

The radius of the near-IR emission, where dust can be sublimated by stellar radiation, can also set the physical radius of the inner boundary of the disk \citep{Dullemond2001,Monnier-Millan2002,Hillen2017,Kluska2019,Lazareff2017}. Using the following luminosity--radius relation \citep[e.g.][]{Lazareff2017}, we estimated the inner-rim radius (R\textsubscript{rim}) of the CBD for U~Mon:
\begin{equation}
\label{equation:innerrim}
R_{\rm rim} = \frac{1}{2}\,\left(\frac{C_{\rm bw}}{\epsilon}\right)^{1/2}\,\left(\frac{L_{\rm \star}}{4\pi\,\sigma\,T_{\rm sub}^4}\right)^{1/2}\,
\end{equation}
where we adopted L$_{\rm \star}$ = 5480~L$_{\rm \odot}$ \citep{Bodi2019} as the stellar luminosity. C$_{\rm bw}$ is the back-warming coefficient of the inner-disk edge that ranges between $\sim$1 and 4, where C$_{\rm bw}=1$ provides a lower limit on R$_{\rm rim}$ and C$_{\rm bw}=4$ provides an upper limit on R$_{\rm rim}$ \citep{Monnier2005,Kama2009}. The cooling efficiency of the dust grains, defined by $\epsilon$ = $\kappa$(T$_{\rm dust})$/$\kappa$(T$_{\rm \star})$, is a ratio between the Planck mean opacity ($\kappa$) of the dust species at its own temperatures and that at the stellar temperature, where $\epsilon \leq 1$, and generally increases with grain size \citep{Kama2009}. For the dust in the CBD, we assume $\epsilon \sim$~1. Typically, for oxygen-rich dust species ([C/O] of 0.8 for U~Mon), the sublimation temperature (T$_{\rm sub}$) is $\sim$1100~K \citep{Bladh2013}. We note that \citet{Kluska2019} suggest a higher inner-dust temperature ($\sim 2600$~K) for U~Mon but only reach a moderate fit with their most complex model, whereas most of their sample of post-AGB binaries have near-IR circumstellar emission sublimation temperatures lower than 1200~K. Finally, the range between C$_{\rm bw} =1$ and C$_{\rm bw} = 4$ gives an R$_{\rm rim}$ between $\sim$4.5 and 9.0~au, respectively.

\subsection{Submillimeter Emission from U~Mon}
\label{ssec:sma_umon}

The SMA observations in the combined array configurations and uniform weighting of visibilities resulted in continuum maps with synthesized beam sizes of 0\farcs69 $\times$ 0\farcs37 at 1.3~mm and 0\farcs39 $\times$ 0\farcs24 at 0.87~mm, respectively. Simple fits of elliptical Gaussians to the map at 0.87~mm give a source size of (0\farcs89 $\pm$ 0\farcs34) $\times$ (0\farcs20 $\pm$ 0\farcs39) and an orientation of the longer axis at a position angle (P.A.) of 55\degr $\pm$ 24\degr. The map at 1.3~mm indicates that the source size is much smaller than the beam, i.e. with FWHM~$\lesssim$~0\farcs55. 
To better constrain the size and make best use of the data, we directly tried to fit a model source to the calibrated visibilities. 
Best solutions for an elliptical Gaussian and an elliptical uniform disk were found in a {\tt CASA} task {\tt uvmodelfit} and are presented in Table~\ref{Tab:sizes}. 

\begin{table}[!ht]
\caption{Results of model fits to visibilities of U~Mon measured with the SMA}
\begin{center}
\begin{tabular}{ccccc}
\hline \noalign {\smallskip}
Band & Model & Major & Minor/Major & PA \\
 & & (mas) & & (deg) \\
\hline \noalign {\smallskip}
1.3~mm & Gaussian   & 299 $\pm$ 11 & 0.73 $\pm$ 0.04 & 83 $\pm$ 5\\ 
1.3~mm & Unif. disk & 483 $\pm$ 24 & 0.76 $\pm$ 0.05 & 84 $\pm$ 8\\
0.8~mm & Gaussian   & 244 $\pm$ 6  & 0.86 $\pm$ 0.03 & 55 $\pm$ 7\\ 
0.8~mm & Unif. disk & 482 $\pm$ 64 & 1.0  $\pm$ 0.2  & $\cdots$    \\ 
\hline
\end{tabular}
\tablecomments{\small For Gaussian models, FWHM is given as the major-axis size.}
\label{Tab:sizes}
\end{center}
\end{table}

The model fits indicate a nearly circular source with a Gaussian FWHM smaller than about 500~mas, which corresponds to a disk diameter $\lesssim$550~au at the nominal distance of 1.1~kpc. A denser {\it uv} coverage of observations at longer baselines would be necessary to better constrain the size of the submillimeter source.

The SMA fluxes are consistent with earlier observations in the nearby bands and represent the coolest component of U~Mon's SED, which is consistent with a dusty source of T\textsubscript{d} = 350~K (see Section~\ref{ssec:results_SED}).

\subsection{X-Ray and {\normalfont UV} Emission from U Mon}

This is the first X-ray detection of an RV~Tauri variable by any X-ray telescope. The X-rays are consistent with the location of U Mon within the pointing uncertainties and the fairly broad point spread function of the XMM EPIC X-ray detectors (FWHM $\sim 12^{\prime\prime}$). There is an additional but fainter source near the position of U~Mon with a separation of $\sim20^{\prime\prime}$ (see Figure~\ref{fig:umon_images}). Given the relative brightnesses of the two sources and the radius used to extract the X-ray products, contamination by the nearby source is minimal. 

\begin{figure}[!ht]
\centering
\includegraphics[height=2.5in]{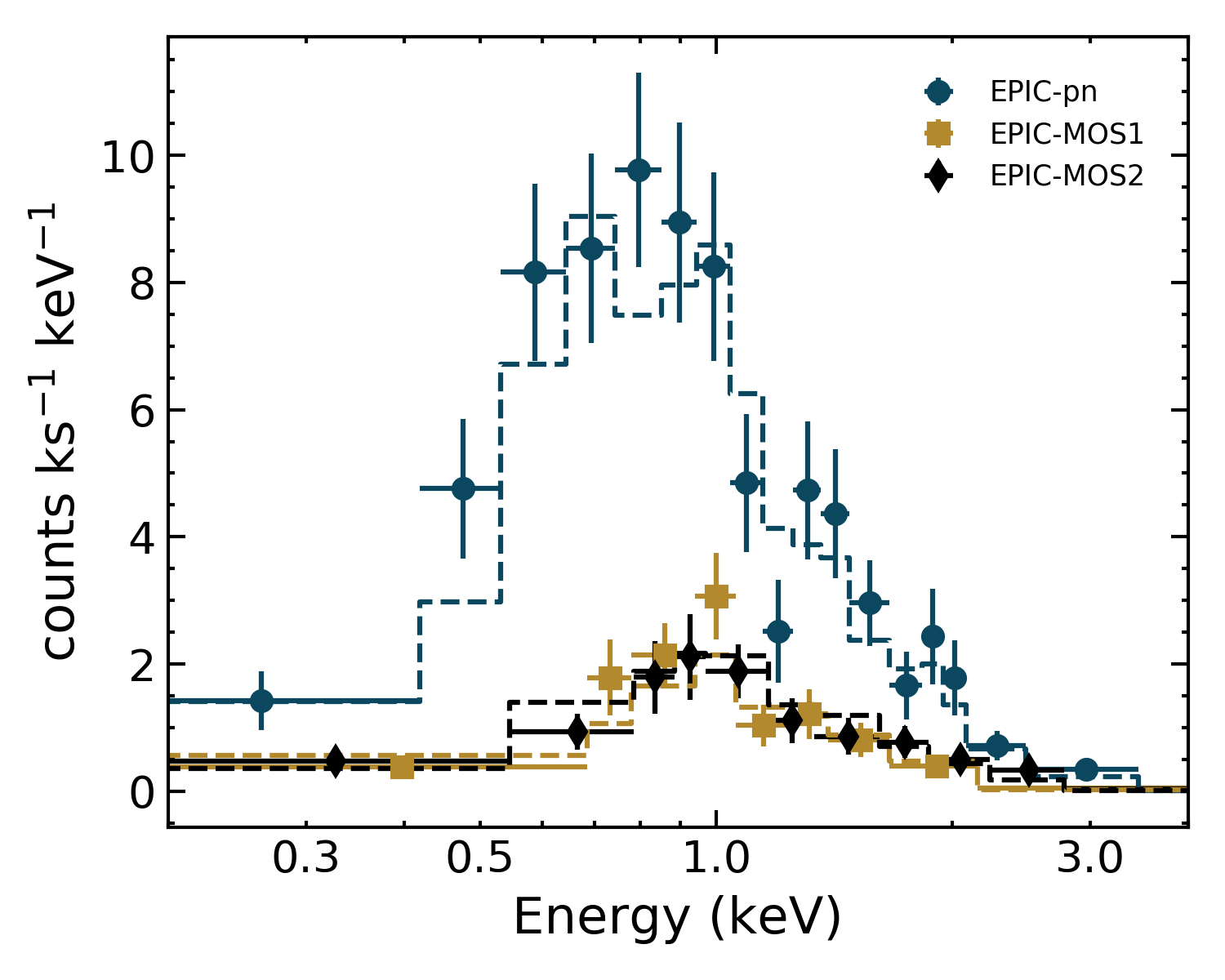}
\caption{\small U~Mon's X-ray spectra and best-fit spectral model (see Table \ref{Tab:xrayfit}).}
\label{fig:xr_spectra}
\end{figure}

\begin{figure}[!ht]
    \centering
    \includegraphics[height=2.5in]{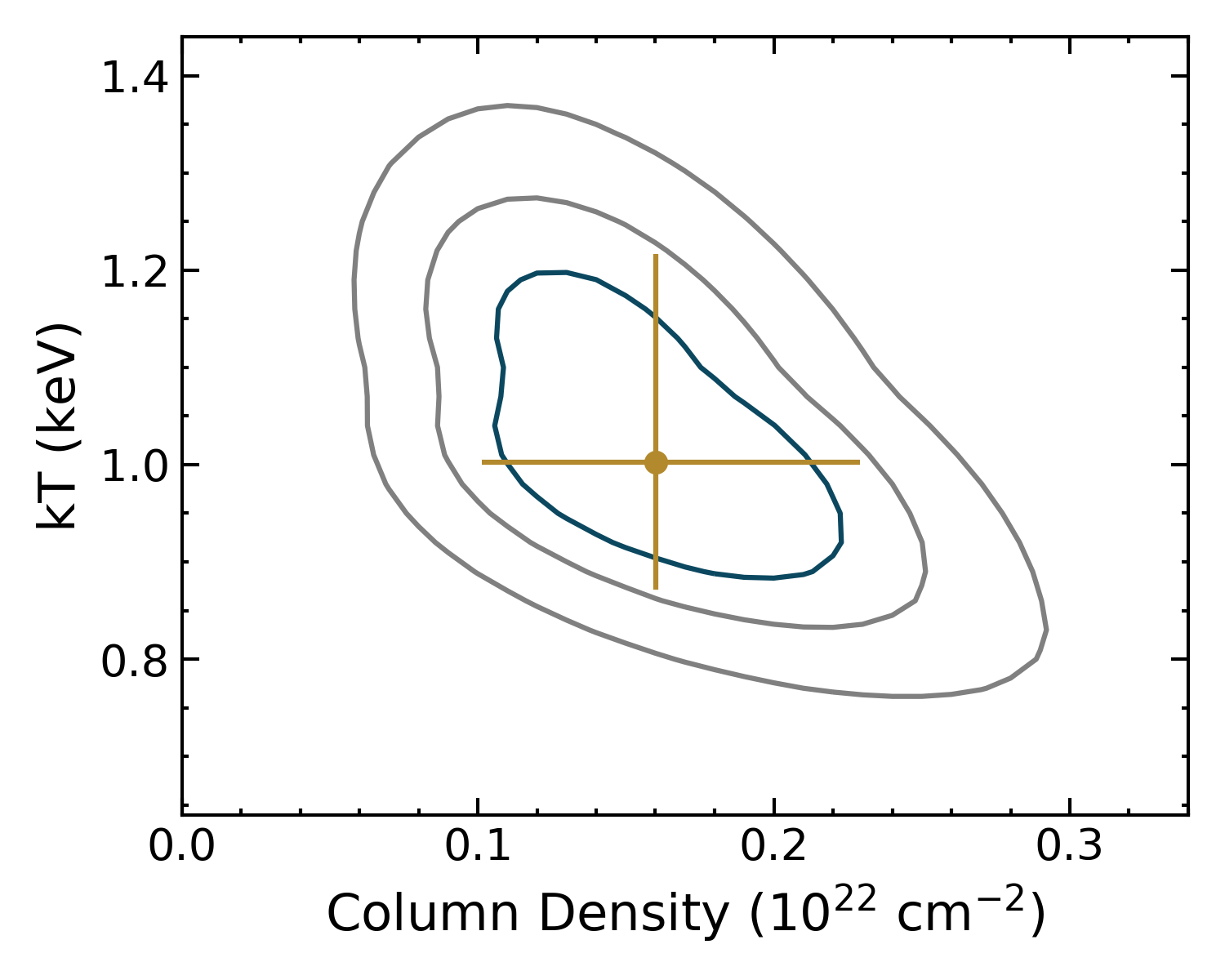}
    \caption{\small Best-fit confidence map for the plasma temperature ($kT$) and column density ($N_{\rm H}$) of the X-ray spectral fit. The best-fit solution is depicted with a 98\% confidence range, and 68\%, 90\%, and 98\% confidence levels are depicted.}
    \label{fig:xspec_contours}
\end{figure}

The spectra were modeled using an absorbed \citep[{\tt tbabs};][]{Wilms2000}, optically thin plasma model with variable abundances \citep[{\tt vapec};][]{Smith2001,Foster2012}. Model fitting was performed with {\tt XSPEC} \citep{Arnaud1996}. No acceptable fit could be found with solar abundances, but adequate fits were found when Fe was allowed to vary. Given the limited energy resolution and moderately low count rate of the CCD spectrum, we could not obtain meaningful constraints on both the plasma properties (temperature and normalization) and the important elemental abundances in the energy range of the detected emission (C, N, O, Ne, Mg, and Fe). Instead, we fixed the Fe abundance to a previously reported value of [Fe/H]~=-0.79 and C-to-O ratio to 0.8 \citep{Giridhar2000ApJ}. The O abundance was allowed to vary along with the overlying absorption, plasma temperature, and plasma model normalization. 
The best-fit parameters are presented in Table~\ref{Tab:xrayfit} along with the absorbed and intrinsic X-ray fluxes and source luminosity for a distance of 1.1~kpc \citep{Bodi2019}. The best-fit spectral model is displayed with the X-ray spectra in Figure~\ref{fig:xr_spectra}, and the unfolded spectrum is presented in the multiwavelength SED in Figure~\ref{fig:UMon_SED}; confidence level contours are shown in Figure~\ref{fig:xspec_contours}. 

\begin{table}[!ht]
\caption{X-ray Spectral Fit Results for U~Mon}
\begin{center}
\begin{tabular}{lc}
\hline \noalign {\smallskip}
Quantity & Value \\
\hline \noalign {\smallskip}
$N_{\rm H}$ ($10^{21} {\rm cm}^2$) & $1.2^{+0.6}_{-0.5}$ \\ 
$T_{\rm X}$ ($10^{6}$ K) & $12^{+1}_{-2}$ \\ 
Oxygen abundance\footnote{Abundances are number relative to solar values from \citet{1989GeCoA..53..197A}. Iron abundance has been fixed to  the [Fe/H] value of \citet{Giridhar2000ApJ} after conversion to the number relative value.} & $3.4^{+3.0}_{-1.6}$ \\ 
Iron abundance & 0.16 \\ 
Model normalization ($10^{-5}$) & $2.1^{+0.3}_{-0.3}$ \\
$F_{\rm X, abs}$\footnote{Model-derived fluxes have been corrected for the encircled energy fraction of the extraction region.} ($10^{-14} {\rm ~erg~cm}^{-2}{\rm ~s}^{-1}$) & $2.2^{+0.4}_{-0.4}$ \\ 
$F_{\rm X, unabs}$ ($10^{-14} {\rm ~erg~cm}^{-2}{\rm ~s}^{-1}$) & $3.4^{+0.6}_{-0.6}$ \\ 
EM ($10^{53}  {\rm ~cm}^{-3}$) & $3^{+1}_{-1}$ \\
$L_{\rm X}$ ($10^{30} {\rm ~erg~s}^{-1}$) & $5^{+1}_{-1}$ \\ 
\hline
\end{tabular}
\label{Tab:xrayfit}
\end{center}
\end{table}

Our analysis of the X-ray observations suggests that the emission is consistent with an iron-deficient hot ($\sim10$~MK) plasma with moderate absorption. The absorption reported in Table~\ref{Tab:xrayfit} is consistent with extinction to U Mon \citep[E$_{B-V} $$\sim$~0.3~mag;][]{Deruyter2005RVTau}. The X-ray light curve reveals no evidence of flaring activity at the 3$\sigma$ level; however, the light curves required 10 ks temporal bins to net sufficient signal-to-noise ratio. Hence we cannot evaluate activity on shorter time-scales. We discuss the origin of the X-ray emission in further detail in Section~\ref{ssec:discussion_xrays}. 

The XMM Optical Monitor performed numerous exposures in three UV filters: UVW1, UVM2, and UVW2 (with effective wavelength of 0.291, 0.231, and 0.212~$\micron$, respectively). U~Mon is detected in all three bands but saturated in UVW1. The UVM2 and UVW2 measurements are included in the multiwavelength SED in Figure~\ref{fig:UMon_SED} and in Table~\ref{Tab:SED}. The UV light curves show variability consistent with Poisson noise. Given the short time frame for the UV observations, we are not very sensitive to variations or trends related to the pulsation period or the long-period variation. 

\section{Discussion}\label{sec:discussion}

\subsection{A 60 yr Trend in the Light Curve}\label{ssec:60yr_trend}

U~Mon's RVb behavior has been observed since before the 1900s, giving a window into the behavior of the CBD. Even with the lack of uniform monitoring in combination with various physical processes present (e.g., pulsations), an even slower trend in the light curve is still detected in the AAVSO+DASCH light curve. 

In particular, three RVb cycles between 1931 and 1951 have significantly more scatter, compared to the overall light curve. They resemble three cycles between 1991 and 2011 (see Figure \ref{fig:lc_phased_60yr}). The disk-shadowing interpretation can be a possible explanation for the disappearance of the long-term variation or the lack thereof. This phenomenon can represent times when the RVb minima have become shallower and shorter, perhaps indicating some kind of precession of the binary or of disk-structure inhomogeneity at the inner-disk edge caused by binary--disk interaction \citep[e.g.][]{Kluska2018,Oomen2020}.

Furthermore, the light curve hints of a different feature of two even deeper RVb minima that occur around 1925 and more clearly around 1985 (phase $\sim$34~yr in Figure \ref{fig:lc_phased_60yr}) with the same time separation of $\sim$60.4 yr. A warp or a dense feature at the inner edge of the disk orbiting around it would only take about 6~yr (2190~days), which is comparable to the orbital period. If a warp or dense feature is farther out in a Keplerian orbit, it would have to be at a distance of 25~au to block out more of the post-AGB so that the RVb minimum becomes deeper every 60.4~yr. 

As more photometric monitoring and interferometric imaging become available for post-AGB binaries, the detection of CBD warps or features may give rise to future investigations on possible second-generation planet formation such as in studies around other evolved stars \citep[e.g.][]{Hardy2016,Homan2018}.

\subsection{Circumbinary Disk Interaction}\label{ssec:discussion_bdi}

\citet{Pollard1995} and \citet{Pollard2006} found enhanced H$\alpha$ emission in U~Mon during the RVb minima. This phenomenon is reminiscent of one observed in other binary star systems with eccentric orbits within CBDs, in which matter is pulled from the inner edge of the disk and onto the stars at apastron \citep[e.g.,][]{ArtymowiczLubow1996,Basri1997}. Based on the periastron time occurring at RVb maximum, we experiment with the idea that the relevant time to explain the interaction that produces H$\alpha$ enhancement is not at periastron but rather the time when the two stars are farthest apart (apastron), which would be when the stars would presumably be most likely to interact with the inner edge of the CBD. 

Based on our orbital fit for the binary, the argument of periastron ($\omega$ = $95^\circ \pm 7^\circ$) means that the longest dimension of the ellipse is oriented parallel to our line of sight. In other words, when the two stars are farthest apart, they are also along our line of sight during apastron, which happens during RVb minima. This scenario requires that the post-AGB be mostly hidden by the CBD while the companion travels in the background at an elevated angle crossing our light of sight and possibly interacting with the inner edge of the CBD, causing the enhanced H$\alpha$ emission observed by \citet{Pollard2006} during RVb minima. This is also shown from the radial velocities in Figure~\ref{fig:orbit}, which measure the motion of the post-AGB star. The orbital phase of zero is at periastron (which corresponds to RVb maximum).

The nature of the CBD--binary interaction is not yet clear in post-AGB binaries; however, processes such as photospheric chemical depletion show that it is crucial. Interestingly, U~Mon's photosphere does not show signs of depletion \citep{Giridhar2005,Gezer2015}; this could be because depletion is mostly observed in stars hotter than 5000~K \citep{Venn2014}. It is also possible that the depletion process could be interrupted by wind Roche lobe overflow onto the companion \citep{Mohamed2007}.  

The recent study by \citet{Oomen2020} investigated a few disk--binary interaction mechanisms for post-AGB binaries with orbital periods of 100--2500 days and eccentricities $\gtrsim0.3$. They concluded that disk--binary interactions are unlikely to pump the eccentricity to the observed values within the evolutionary timescales for the stars they sampled. U~Mon is not very eccentric (e = 0.31); this may suggest a weak disk--binary interaction.

However, the \citet{Kluska2019} geometric image reconstruction model of U~Mon's visibility profile resulted in a fit that displayed strong azimuthal modulation, showing that this system likely has a very complex inner-disk morphology. Additionally, they showed that U~Mon has a relatively large mid-IR-to-near-IR size ratio. The authors note that it is possible that the model is not able to reproduce all the complexities, which could have an effect of on their derived size of the CBD's inner rim; thus, further observations will bring stronger constraints. Such complexities have been seen in some YSO CBD systems, which can be regarded as scaled-down versions of the CBDs around post-AGB binaries \citep[e.g.][]{Hillen2017}. For example, in the CBD of the young binary GW~Ori, imaging observations by \citet{Kraus2020} find evidence for an inner ring and a large warp in the outer part of the disk. This phenomenon occurs when the outer disk is misaligned with the orbital plane, so it wraps and breaks into precessing rings, which may provide a mechanism for planet formation.

\citet{Bollen2019} provide a detailed investigation on the possibilities for jets in explaining the H$\alpha$ emission feature observed in post-AGB systems. In their model, the H$\alpha$ feature is superimposed on the photospheric absorption from the post-AGB star and is present throughout the entire orbit. To establish whether such a jet scenario may apply to U~Mon as well would require phase-resolved, high-resolution spectroscopy of the H$\alpha$ feature.

If, instead, there is a stream outflow coming off the CBD onto the companion, always traveling around the companion and opposite of the post-AGB primary, the H$\alpha$ enhancement would have a timescale near the binary orbital period. The stream of material would be at a focus point where it would be enhanced around apastron because that would be the shortest distance the CBD material would flow onto the companion. The stability for the shortest orbit of material around a binary should be at least $\sim$3--4 times the binary separation \citep{HolmanWiegert1999}. However, it is possible that the inner edge of RV~Tauri disks may not be in long-term stable orbits. \citet{ArtymowiczLubow1996} showed hydrodynamical models for unstable inner-disk rims, so it is possible that material may be stripped by the binary, in particular the companion. This scenario may also explain the X-rays observed in U~Mon. 

\subsection{Nature and Origin of X-Ray Emission from U~Mon}\label{ssec:discussion_xrays}

As a binary system with a CBD, there are a number of processes that can generate the detected X-ray emission. We consider processes from U~Mon, the companion, and interactions between the components of the system including the CBD.

If the X-ray emission originates from U Mon, the X-ray luminosity corresponds to an $L_{\rm X}/L_{\rm bol} = 10^{-7}$. Generally, such a ratio is consistent with the X-ray emission that arises from shocks caused by variable winds in O and B stars; however, the X-ray emission from U Mon does not quite follow the general trend with L$_{\rm bol}$ seen for B stars in the Carina Nebula, which would predict an X-ray luminosity up to an order of magnitude higher \citep[e.g.][]{Naze2011}. The plasma temperature ($>$10~MK) is consistent with hot components found in the Carina O and B stars; however, no fast outflows have been measured from U Mon. Indeed, the large radius of U Mon suggests an escape velocity $<$100~km~s$^{-1}$, whereas the plasma temperature requires velocities $>$800~km~s$^{-1}$ for a strong shock. Such velocities have not been measured from U Mon and are unlikely to be produced by any pulsation-induced shocks in the atmosphere of U Mon \citep{Fokin1994}, unlike that suggested for X-ray emission from the Cepheid star $\delta$ Cep \citep{2017ApJ...838...67E}. \citet{Gillet1990} measured shock amplitudes from spectroscopy for the RV~Tauri variables R~Sct and AC~Her that did not exceed $\sim$40 km s$^{-1}$.

We note that \citet{Moschou2020} recently investigated a mechanism to produce phase-dependent shocked gas X-ray emission in the pulsating atmosphere of classical Cepheids. This mechanism requires the presence of solar-like coronal plasma into which the phase-dependent shocks are driven \citep{Moschou2020}. The detected X-ray emission from U~Mon is hotter than that studied by \citet{Moschou2020}, and we lack adequate phase coverage to establish the presence of such shocks from U~Mon, as well as the necessary signal-to-noise ratio in our XMM X-ray light curve to establish the presence of such shocks from U Mon. A solar-like coronal plasma is an intriguing notion given the detection of a magnetic field (very likely dynamo generated) at the surface of this cool post-AGB star, but there is, so far, no information on any related activity such as flares, stellar spots, or dynamo variability that would be linked to X-ray emission. Furthermore, the ability to sustain a hot stellar corona on such a large evolved star is unclear \citep{Sahai2015}.

Accretion, from a compact disk or infalling material from the large CBD (as discussed in Section~\ref{ssec:discussion_bdi}), might create shocks that can heat material up to X-ray-emitting temperatures. However, given the stellar properties of U~Mon in Table~\ref{Tab:Properties}, it is unlikely that infalling material will reach sufficiently high velocities ($<$100~${\rm km}\ {\rm s}^{-1}$ at the surface of U Mon). If the companion star has a radius $\leq$2\ R$_{\odot}$, which would make the companion consistent with an A-type main-sequence star, infalling material, from the CBD or donated by U~Mon via wind Roche lobe overflow \citep{Mohamed2007}, can reach speeds in excess of $800\ {\rm km}\ {\rm s}^{-1}$. Such infalling material can either slam into the surface of the star along magnetic fields or form into an accretion disk around the companion.

An accretion disk forming around the companion could then generate outflows as seen in other post-AGB systems \citep{Gorlova2012_AAP_timespec_jet_bd46442,Gorlova2015}; however, such outflows have yet to be observed from the U~Mon system. If such an outflow exists, its velocity could be sufficient to generate shocks capable of explaining the detected X-ray emission. Detailed monitoring of the H$\alpha$ emission from U~Mon, like that performed of similar systems by \citet{Gorlova2012_AAP_timespec_jet_bd46442,Gorlova2013}, could help establish the presence of accretion-disk-driven outflows.

If the binary companion is the source of X-ray emission, in addition to accretion, coronal activity and radiative stellar winds can produce X-ray emission similar to that given in Table~\ref{Tab:xrayfit}. However, at $\sim$2.2~M$_{\odot}$ (Table~\ref{Tab:Properties}), a main-sequence companion would be consistent with an early-type A star and unable to support magnetic activity to produce coronal X-ray emission \citep{Stelzer_2006}. Furthermore, the bolometric luminosity of such a main-sequence companion is 100 times fainter than U Mon, leading to an $L_{\rm X}/L_{\rm bol}$ of $10^{-5}$ for the companion, which is an unusually high fraction for an A-type star. Such a high fraction of X-ray emission and a high-temperature plasma are comparable to what is seen from Herbig Ae/Be stars, which are intermediate-mass pre-main-sequence stars embedded in dusty disks \citep{Stelzer_2006,Stelzer_2008}. In the case of Herbig stars, the high plasma temperature suggests that radiative winds are not the emission mechanism, and higher-order binary components, namely, unseen rapidly rotating late-type companions to the A star in a Herbig system, have been suggested as a potential origin for their X-ray emission \citep{Stelzer_2006}. Since we have not observed the companion and only have a modest constraint on its mass ($2.2^{+1.0}_{-0.75}$~M$_\odot$; see Table~\ref{Tab:Properties}), we cannot exclude the remote possibility that the companion is a compact binary. In such a scenario, the detected X-ray emission would be consistent with enhanced coronal activity from rapidly rotating late-type stars in a close binary system. 

We have highlighted a number of potential origins for the X-ray emission from the U~Mon system, each with strong implications for the nature and evolution of the system. Additional supporting information, such as further constraints on the companion properties and monitoring of the H$\alpha$, like that reported in \citet{Bollen2019}, and additional observations of the X-ray emission through the phases of the U~Mon pulsation and orbit, is essential to better understand the origin and influence of the X-ray emission.

\section{Conclusions}\label{sec:conclusions}

We have conducted the most comprehensive characterization yet of an RVb system, U~Mon, with observations spanning the largest range of wavelengths (X-ray to millimeter) and the longest range of time ($\sim$130~yr).

U~Mon, one of the brightest RV~Tauri variables of its class, is shown here to be a binary system comprising a 2.07~M$_\odot$ post-AGB star that has lost mass relative to a more massive unevolved 2.2~M$_\odot$ A-star companion, and is surrounded by a large CBD. We obtained new orbital parameters within the errors of the values found by \citet{Oomen2018}, by fixing the orbital period as the photometric RVb period. The new periastron value, T\textsubscript{0}~=~$2452203$~days, reveals that the apastron times align with RVb minima (Figure~\ref{fig:orbit}). Additionally, based on our orbital fit, the argument of periastron ($\omega$~=~$95^{\circ}$) translates to the long axis of the orbit being roughly aligned along our line of sight. This supports the argument that apastron occurs when the post-AGB star is occulted by the near side of the disk at RVb minimum while the A-type star companion is most revealed at the far side of the inner hole in the CBD. Moreover, we found that the binary's semimajor-axis separation is $5.78$~au, which at apastron is comparable to the size range of the inner-disk hole radius ($\sim$4.5--9.0~au).

U~Mon shows several interesting features, some not yet seen in any other post-AGB binary, such as a magnetic field and most recently X-rays. We found a $\sim$10~MK plasma model that is consistent with the X-ray spectra and considered origins from U~Mon and/or its companion. We suggest that at apastron the companion may be close enough to strip material from the CBD, creating the X-ray emission observed. This interpretation may agree with the complex morphology found in the interferometric dataset for U~Mon by \citet{Kluska2019}.

The occurrence of the enhanced H$\alpha$ at RVb minima \citep[e.g.][]{Pollard2006} also seems to correspond to when the A-type star companion is most revealed at apastron, implying that since the post-AGB star is shadowed by the CBD at this phase, the enhanced H$\alpha$ must not be coming from the post-AGB star but rather from the environment around the A-type companion, likely an accretion disk \citep[e.g.][]{Bollen2019}. Recent work has shown that the H$\alpha$ variation is due to a bipolar jet from the circumcompanion accretion disks in other post-AGB binaries. More time series of optical spectra are needed to investigate the H$\alpha$ variability in U~Mon.

We tested whether additional blackbody components with properties of the A-type star companion would influence U~Mon's SED but found that such a companion seems to only be detected in H$\alpha$ (and possibly X-rays), even at RVb minimum, when it is most revealed in its orbit. The X-ray through millimeter emission, however, is consistent with components that include a hot thermal plasma, the stellar photosphere of the post-AGB, and the inner- and extended-disk emission. The scaling of the modified blackbody for the dust emission includes parameters such as opacity and dust grain sizes that we do not explore in this paper. However, U~Mon's new SMA observation fluxes are in agreement with the slope of the Rayleigh-Jeans form of the Planck function, between the other 60--3000 $\mu$m flux measurements. Such slopes are consistent with the presence of a composition of large dust grains \citep[radius $\gtrsim$ 0.1 mm;][]{Deruyter2005RVTau}. Due to the short SMA observations, we are only able to estimate upper limits on U~Mon's disk diameter size $\lesssim$550~au (see Table \ref{Tab:sizes}).

Nevertheless, with the AAVSO+DASCH light curve, which spans more than 130~yr, we find evidence for a 60.4 yr cycle that could be due to some structure in the disk at a separation of $\sim$25~au that temporarily allows the post-AGB star to avoid being occulted at apastron/RVb minimum. Accordingly, the 60.4 yr cycle also produces an especially deeper RVb minimum (phase $\sim$54~yr in Figure \ref{fig:lc_phased_60yr}--two cycles before the ``disappearance'' of the RVb minima). Presumably this can be a corresponding feature in the disk that causes an exceptionally large occultation of the post-AGB star. More radio observations at longer baselines would constrain the size of the disk, as well as resolve features we see in U~Mon's light curve.

Furthermore, the detection of X-rays from the U~Mon system has opened up new possibilities; do all RV~Tauri variables exhibit X-ray emission? In order to answer this question, X-rays will be an important factor to consider in forthcoming RV~Tauri studies to enable constraints on their evolution \citep[e.g.,][]{Graber_Montez_2021}.

Most importantly, U~Mon now becomes an archetype for the study of binary post-AGB stellar environments that represent an important evolutionary phase, which either sets the stage for sculpting the morphology and evolution of planetary nebulae or which may represent systems that never become planetary nebulae at all, as the disk and stellar companion arrest its development.

\acknowledgments{
We acknowledge the very helpful review of the anonymous referee, which substantively improved the paper. 
We acknowledge the excellent work of the NASA's Goddard Space Flight Center's Astrophysics Science Division Communications Team in the production of a NASA news feature for this research.
L.D.V. and K.G.S acknowledge the support of the NASA MUREP Harriett G. Jenkins Predoctoral Fellowship, Grant Nos. NNX15AU33H and 80NSSC19K1292.
R.M.Jr. and L.D.V. acknowledge the federal support from the Latino Initiatives Pool, administered by the Smithsonian Latino Center.
This research has made use of XMM-Newton, an ESA Science Mission with instruments and contributions directly funded by ESA Member States and the USA (NASA). 
This research has made use of the Submillimeter Array, a joint project between the Smithsonian Astrophysical Observatory and the Academia Sinica Institute of Astronomy and Astrophysics, and is funded by the Smithsonian Institution and the Academia Sinica.
We acknowledge with thanks the variable star observations from the AAVSO International Database contributed by observers worldwide and used in this research. 
This research has made use of the DASCH project at Harvard, which is grateful for partial support from NSF grants AST-0407380, AST-0909073, and AST-1313370. 
This research has made use of the SIMBAD database, operated at CDS, Strasbourg, France, and of the VizieR catalog access tool, CDS, Strasbourg, France (\href{https://doi.org/10.26093/cds/vizier}{{DOI: 10.26093/cds/vizier}}). The original description of the VizieR service was published in 2000 \citep{Ochsenbein2000}.}

\vspace{5mm}
\facilities{AAVSO, DASCH, SMA, XMM-Newton}

\software{
{\tt Astropy} \citep{Astropy2013}, 
{\tt CASA} \citep{CASA}, 
{\tt MIR} ({\url{http://github.com/qi-molecules/sma-mir}}), 
{\tt PHOEBE} \citep{Prsa2016}, 
{\tt XSPEC} \citep{Arnaud1996}}

\bibliography{main.bib}

\end{document}